\documentstyle[11pt,newpasp,psfig]{article}
\def\ee #1 {\times 10^{#1}}
\def\ut #1 #2 { \, \hbox{#1}^{#2}}
\def\u #1 { \, \hbox{#1}}

\def\msol{\, \hbox{$\hbox{M}_\odot$}}

\let\grad=\nabla
\def\cross{{\bf \times}}
\def\curl #1 {\grad \cross #1}
\def\div #1 {\grad \cdot #1}

\begin{document}

\title{The Nature of  3EG J1746-2851
 in the Nucleus of the Galaxy}
\author{F. Yusef-Zadeh}
\affil{Dept of Physics and Astronomy, Northwestern University,
Evanston, IL. 60208}
%\author{Winston Churchill}
%\affil{Imperial College London, Blackett Laboratory, Prince Consort
%Road, London SW7 2BW}

\setcounter{page}{1}
% this page number will be filled later by the editors....
\index{de Gaulle, C.}
\index{Churchill, W.}

\begin{abstract}
There are a number of candidate sources that 
are considered  to be responsible for the origin 
of the EGRET source 3EG J1746--2851 (Hartman
et al. 1999).  The lack of clear identification of this source 
with any known sources in this region is due
to the
complexity and
the
richness of the environment
of the Galactic center as well as the low spatial resolution of the
EGRET
observations.
Four hypotheses,  which include the interaction of something with
something else,  are described
to account for the energetics and the spectrum of 
the $\gamma$-ray emission from the enigmatic EGRET source which lies
within
0.2$^0$ of the Galactic center.  
The possible interaction sites result from the following: i) the
nonthermal filaments of
the radio Arc
\&
G0.13-0.13 molecular cloud, ii) the supernova remnant Sgr A East \&
the 50
km s$^{-1}$ molecular cloud, iii) the magnetized
filaments of the Arc \&
M0.20-0.033
molecular cloud and iv) the relativistic particles of the Arches
cluster \&  its  own stellar radiations field.

\end{abstract}

\section{Introduction}

In a recent paper, Yusef-Zadeh, Law and Wardle (2002)  argued that
the molecular cloud G0.13--0.13 is interacting with the nonthermal
radio filaments of the Arc (Yusef-Zadeh, Morris \& Chance 1984) and
is responsible for the enhanced 6.4 keV line emission at the
interface between these two features.  This argument was also used in
support of an earlier interpretation by Tsuboi et al. (1997) and Oka
et al.  (1997) who showed the morphological and dynamical evidence
for such an interaction.  Figure 1 shows  radio continuum image of the 
straight nonthermal filaments (red color), the G0.13--0.13 cloud
(contours) 
and the diffuse and filamentary
X-ray features (green color) lie within the 95\% error circle of an
unidentified
EGRET source 3EG J1746--2851 (blue color)
 (Hartman et al. 1999), with
a steady
source of strong $\gamma$-ray emission between 30 MeV and 10 GeV.  
The photon index is $\alpha$=1.7$\pm0.7$, and its flux is estimated
to be 1.2$\times10^{-6}$ photons cm$^{-2}$ s$^{-1}$ with energies
greater than 100 MeV, corresponding to $10^{40}$ photons s$^{-1}$ at
the distance of the Galactic center (Hartman et al. 1999).

3EG J1746--2851 is interpreted in terms of the interaction of the
nonthermal filaments with the molecular cloud.  The $E^{-1.7}$ photon
spectrum of 3EG J1746--2851 matches onto the cloud spectrum at about
10 keV
when extended down to X-ray energies.  This suggests that the EGRET
source
is produced by\\
 bremsstrahlung, and that the electron spectrum
extends up to GeV energies  with an
$E^{-1.7}$ dependence in this range.
This model is supported by the spectral index  
measurements of the nonthermal filaments between cm and mm
wavelengths.  The spectral indices ($p$, where $S_\nu \propto \nu^p$)
are positive along the vertical filaments of the Arc but there is an
anomalous filament near G0.16-0.15 with $p\sim-0.35$ (Anantharamaiah
et al.  1992), consistent with synchrotron emission from GeV
electrons  
with an $E^{-1.7}$ spectrum.

Another interpretation of the EGRET source  considers
that the relativistic particles of the nonthermal
supernova remnant Sgr A East near  the Galactic center
is interacting with the 50 km s$^{-1}$ molecular cloud
(Melia et al.
1998). The bright oval-shaped structure to the southwest
of Figure 1 shows Sgr A East.  The black spot drawn on 
Sgr A East coincides with Sgr A$^*$, the massive black hole
at the Galactic center. 
% (Eckart and Genzel 1997; Ghez et al. 1998). 
The p-p scattering resulting from this interaction
generates pions  which eventually decay into $\gamma$-rays.
These authors  show that the electron-positrons produced from pion
decays
fit the radio spectrum of Sgr A East well.
Both Sgr A East and Sgr A$^*$ exclude 
the 95\% error circle of the  EGRET source. 
However, 
the EGRET source could include  Sgr A East if the $\gamma$-ray
emission is extended.

Another scenario involves  the possibility  
that  the EGRET source results from the motion of 
a molecular
cloud M0.2--0.033
pushing against the flux tubes of the nonthermal
radio filaments of the Arc (Pohl 1997). The HII region G0.18-0.04 is
at
the interaction zone  between these two features. The relativistic  
particles of the nonthermal filaments with a monoenergetic electron     
distribution are then responsible for
Compton scattering of far-IR photons of the molecular cloud
G0.2--0.033.

Lastly, the source of $\gamma$-ray emission could arise from
the Arches cluster  (e.g. Blum et al. 2001)
which lies within the 95\%
error circle of 3EG J1746-2851. The remarkable cluster is embedded
within 
an HII complex  and its location is drawn as a cross in Figure 1. 
This cluster 
consists of mainly 150 O star candidates with stellar masses greater
than
20~M$_\odot$.  The Arches cluster is $\sim15''$ across, with an
estimated
density of $3\times10^5$ \msol pc$^{-3}$ within the inner 9$''$ (0.36
pc)
of
the cluster (e.g. Cotera et al. 1996; Serabyn, Shupe \& Figer 1998;
Blum  
et al. 2001).
The ensemble of
colliding winds within the cluster will produce shock waves that
can accelerate particles to high energies. The relativistic particles
of the cluster can then be used to produce $\gamma$-ray
emission by
scattering of the IR radiation field of the cluster. A more detailed   
account  of this proposal will be given elsewhere.

\bigskip

\begin{figure}[t]
%\centerline{
%\psfig{figure=f1.eps,width=11cm,angle=-90}
%\psfig{figure=fig1.ps,width=11cm}
%}
\caption{ A $\lambda$20 cm continuum image of the nonthermal Sgr A
East to the southwest 
and the filamentary Arc  to the northeast is shown in red. The black
spot drawn 
near 
the center of the Sgr A East shell  coincides with the position of
Sgr A$^*$, the massive black hole at the Galactic center.  The cross
indicates the position of the Arches cluster which is
surrounded by curved ionized features known as the arched
filaments. 
The distribution of 3EGJ1746-2851 
is shown in blue. The radius of the 95\% error circle is 0.11$^0$
(Hartman et
al. 1999).  The distribution of X-ray emission near 
the filaments of the Arc is shown in green whereas the 
distribution of CS (1--0) molecular line emission is 
represented  as yellow contours. Note that the distribution of
molecular gas 
and X-ray emission from the regions of Sgr A East and the arches 
cluster are not shown in this figure. (The color version of this
figure  is available in astroph)}
\label{fig1}
\end{figure}

In conclusions, we believe an understanding of the origin of the
EGRET source toward the Galactic center yields insights into the
physical processes occuring in the interaction sites described
here.  There may be interesting aspects of this interaction
picture that can be applied to the nuclei of
normal galaxies or  AGNs 
even if the $\gamma$-ray emission is not produced 
by the massive
black
hole at the dynamical center of the Galaxy. 
Future $\gamma$-ray
observations of this region with Integral
should be useful  to clarify the nature of the enigmatic source
3EG J1746-2851.

%\acknowledgements
% We are very grateful 

\end{document}